\documentclass[%
 reprint,
superscriptaddress,
 amsmath,amssymb,
prb,
]{revtex4-2}

\usepackage{amsmath}
\usepackage{siunitx}
\usepackage{dcolumn}
\usepackage{hyperref}
\usepackage{textcomp}
\usepackage[dvipsnames]{xcolor}
\usepackage{bm}
\usepackage{graphicx}

\usepackage{orcidlink}
\usepackage{fontawesome5}

\preprint{APS/123-QED}

\begin{document}

\title{Unbiased Random Number Generation using Injection-Locked Spin-Torque Nano-Oscillators}

\author{Nhat-Tan Phan\,\orcidlink{0000-0002-2806-1660}} 
\affiliation{Univ. Grenoble Alpes, CEA, CNRS, Grenoble INP, SPINTEC, 38000 Grenoble, France}

\author{Nitin Prasad\,\orcidlink{0000-0001-8088-0164}}
\affiliation{Associate, Physical Measurement Laboratory, National Institute of Standards and Technology, Gaithersburg, Maryland 20899, USA} 
\affiliation{Department of Chemistry and Biochemistry, University of Maryland, College Park, Maryland 20742, USA}

\author{Abderrazak Hakam\,\orcidlink{0009-0006-6555-8840}} 
\affiliation{Univ. Grenoble Alpes, CEA, CNRS, Grenoble INP, SPINTEC, 38000 Grenoble, France}

\author{Ahmed Sidi El Valli\,\orcidlink{0000-0003-1106-2664}} 
\affiliation{Univ. Grenoble Alpes, CEA, CNRS, Grenoble INP, SPINTEC, 38000 Grenoble, France}

\author{Lorena Anghel\,\orcidlink{0000-0001-9569-0072}} 
\affiliation{Univ. Grenoble Alpes, CEA, CNRS, Grenoble INP, SPINTEC, 38000 Grenoble, France}

\author{Luana Benetti\,\orcidlink{0000-0003-3063-957X}}
\affiliation{International Iberian Nanotechnology Laboratory (INL), 4715-31 Braga, Portugal}

\author{Advait Madhavan\,\orcidlink{0000-0002-4121-1336}}
\affiliation{Associate, Physical Measurement Laboratory, National Institute of Standards and Technology, Gaithersburg, Maryland 20899, USA} 
\affiliation{Institute for Research in Electronics and Applied Physics, University of Maryland, College Park, Maryland 20742, USA}

\author{Alex S. Jenkins\,\orcidlink{0000-0002-6188-7755}}
\affiliation{International Iberian Nanotechnology Laboratory (INL), 4715-31 Braga, Portugal}

\author{Ricardo Ferreira\,\orcidlink{0000-0003-0953-2225}}
\affiliation{International Iberian Nanotechnology Laboratory (INL), 4715-31 Braga, Portugal}

\author{Mark D. Stiles\,\orcidlink{0000-0001-8238-4156}} 
 \affiliation{Physical Measurement Laboratory, National Institute of Standards and Technology, Gaithersburg, Maryland 20899, USA}

\author{Ursula Ebels\,\orcidlink{0000-0001-5061-5538}} 
\affiliation{Univ. Grenoble Alpes, CEA, CNRS, Grenoble INP, SPINTEC, 38000 Grenoble, France}

\author{Philippe Talatchian\,\orcidlink{0000-0003-2034-6140}}
\email{philippe.talatchian@cea.fr} 
\affiliation{Univ. Grenoble Alpes, CEA, CNRS, Grenoble INP, SPINTEC, 38000 Grenoble, France}

\begin{abstract}
Unbiased sources of true randomness are critical for the successful deployment of stochastic unconventional computing schemes and encryption applications in hardware. Leveraging nanoscale thermal magnetization fluctuations provides an efficient and almost cost-free means of generating truly random bitstreams, distinguishing them from predictable pseudo-random sequences. However, existing approaches that aim to achieve randomness often suffer from bias, leading to significant deviations from equal fractions of 0 and 1 in the bitstreams and compromising their inherent unpredictability. This study presents a hardware approach that capitalizes on the intrinsic balance of phase noise in an oscillator injection locked at twice its natural frequency, leveraging the stability of this naturally balanced physical system. We demonstrate the successful generation of unbiased and truly random bitstreams through extensive experimentation. Our numerical simulations exhibit excellent agreement with the experimental results, confirming the robustness and viability of our approach.
\end{abstract}

\maketitle

\section{Introduction}
Spin-torque nano-oscillators (STNOs) have emerged as compelling candidates for non-linear, low-power, tunable microwave oscillators \cite{kiselev_microwave_2003, rippard_direct-current_2004}. They hold promise for diverse applications such as broadband microwave communication \cite{choi2014spin,litvinenko2022ultrafast,zhu2023nonlinear} and unconventional computing schemes \cite{grollier2020neuromorphic, finocchio2021promise}. These nanoscale devices, when based on magnetic tunnel junction structures, resemble the memory cells used in magnetic non-volatile memories, enabling straightforward integration with complementary-metal-oxide-semiconductor (CMOS) technology and facilitating their integration into microelectronic chips.

Applying an electrical current to STNOs gives a spin polarized current across the tunnel barrier, generating a spin-transfer torque \cite{berger, slonczewski, ralph2008spin} acting on the magnetization. Above a critical current density, this torque induces self-sustained oscillations of the free-layer magnetization, producing microwave voltage oscillations via magnetoresistive effects. Notably, the frequency of STNOs can be finely tuned by varying the applied direct current or changing applied magnetic fields \cite{slavin_nonlinear_2009}. Within the auto-oscillation regime, STNOs can injection lock \cite{rippard2005injection,quinsat2011injection, dussaux2011phase, hamadeh2014perfect} to external microwave signals, synchronizing \cite{pikovsky2002synchronization} their oscillations to those signals over a frequency span referred to as the injection locking range.

While injection locking has been extensively utilized to reduce phase noise in synchronized STNOs \cite{tamaru2015extremely, kreissig2017vortex, wittrock2021stabilization}, the coherence of magnetization oscillations is susceptible to thermal fluctuations at room temperature, resulting in dephasing events referred to as phase slips \cite{martins2023second}. These phase slips compromise the overall coherence of the system and may impact the performance of various applications. However, in contrast to  conventional efforts to reduce noise, our work aims to harness and exploit the stochastic nature of magnetization dynamics within STNOs. Specifically, we propose a methodology to construct unbiased random number generators at the nanoscale by leveraging the intrinsic phase noise of STNOs.

In the context of encryption \cite{ayubi2020deterministic, abd2020controlled, acosta2017embedded, fischer2012closer} and probabilistic computing schemes \cite{borders2019integer, daniels2020energy, sengupta2016probabilistic, alaghi2018promise}, the generation of high-quality random numbers is paramount. In some applications, pseudorandom bitstreams cannot meet the desired level of unpredictability, as they are periodic, albeit with long periods, and correlated. Several groups have explored spintronic devices, including superparamagnetic tunnel junctions \cite{vodenicarevic2017low, becle2022fast}, and spin-torque nano-oscillators \cite{jenkins2019nanoscale} to overcome this limitation as potential low-energy random number generators. However, these devices often produce stochastic bitstreams that exhibit bias, deviating from an equal probability distribution between the \texttt{0} and \texttt{1} bits, making them vulnerable to malicious actors that can anticipate or predict the next coming bit. Rectifying such biases necessitates the inclusion of additional circuitry, resulting in energy and area cost overheads.

In this paper, we introduce a straightforward approach for unbiased random number generation using naturally unbiased phase states of spin-torque nano-oscillators. By harnessing the intrinsic stochastic phase dynamics of these nano-oscillators under injection locking, we achieve random number generation that is unbiased and truly random based on our statistical analysis. 

Section~II introduces the concept of leveraging the stochastic phase dynamics of STNOs for unbiased random number generation. It describes the phase dynamics of injection-locked STNOs subjected to thermal noise, and how the thermally-induced phase dynamics can be harnessed for generating truly random numbers. This section also presents the experimental demonstration of unbiased random number generation and the specially designed circuit for electrical phase readout, which enables the measurement and characterization of the stochastic bitstream. Building upon this understanding, Section~III focuses on the dwell time tunability of unbiased random number generation via the frequency detuning. It investigates the ability to tune the mean bit flip rate of the generated bit making up the random number bitstream while ensuring the bitsteam is unbiased within the injection locking range. Section~IV statistically tests the generated random numbers. We summarize the essential findings and highlight the significance of unbiased stochastic building blocks for encryption and unconventional computing schemes in the concluding Sec.~V. Appendices A, B, C, and D provide additional details on the vortex-based STNO structure, the free-running and synchronization dynamics of STNOs, the techniques to evaluate the mean dwell times, and the numerical simulation methods that support our findings, respectively.

\section{Leveraging Stochastic Phase Dynamics in Spin-Torque Nano-Oscillators}
\label{section II}

Previous studies on the stochastic phase dynamics in STNOs have primarily focused on enhancing the oscillator's phase coherence, often overlooking the potential of exploiting its inherent randomness. One common technique for improving phase coherence is injection-locking, which involves synchronizing the STNO with a more coherent signal. This synchronization can be achieved by applying an external radio frequency (RF) current at multiple integer frequencies or an integer fraction of the oscillator frequency \cite{LebrunPRL}. In contrast, our approach aims to leverage these injection-locking methods to discretize the inherent phase noise fluctuations of STNOs and generate a stochastic voltage signal that fluctuates between discrete voltage levels. This type of signal is particularly desirable for the binary encoding of a given random number. 

While our primary focus is on random number generation, the implications of our approach extend beyond this application and encompass broader concepts related to STNOs and their phase encoding capabilities. This study serves as a foundation for future explorations of STNO networks that leverage the stochastic phase dynamics for various cognitive computing schemes, such as Ising-based models and microwave encryption applications. Harnessing the inherent randomness of STNOs holds promise for unlocking new phase-based functionalities and applications of STNOs.

\begin{figure}
 \centering
  \includegraphics[width=86mm\centering]{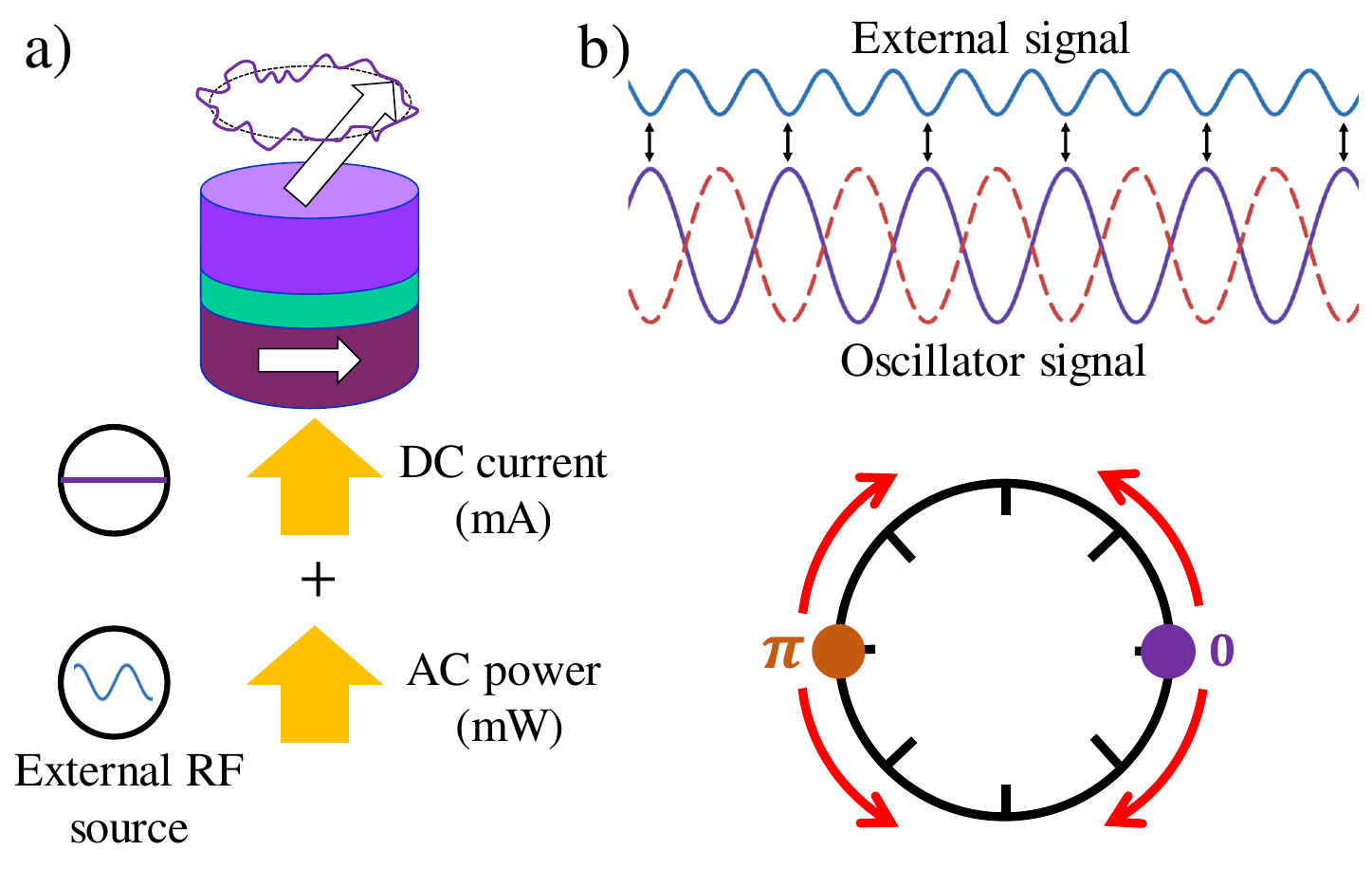}
  \caption{a) Schematic illustrating the use of stochastic phase dynamics of STNOs. A constant DC current $I_\textrm{dc}$ and a fixed out-of-plane magnetic field $H_\textrm{app}$ drive the free layer magnetization into an auto-oscillation regime at a frequency $f_0$. b) We inject an oscillating AC current $I_\textrm{ac}$ with frequency $f_\mathrm{ext}=2f_0$, twice free-running STNO's frequency, leading to synchronization between the STNO's oscillations and $I_\textrm{ac}$. Random fluctuations caused by thermal noise result in a stochastic fluctuation of the phase difference $\Delta\phi$ between in-phase (0) and out-of-phase ($\pi$) states when the STNO is synchronized at close to twice its free-running frequency. To visualize the two states, they are shown on a phasor diagram defined in the reference frame of the STNO.}
  \label{fig1}
\end{figure}

Figure~\ref{fig1} illustrates our approach to leverage the stochastic phase dynamics of STNOs. A constant DC current $I_\textrm{dc}$ and a fixed magnetic field $H_\textrm{app}$ are applied to the magnetic tunnel junction, destabilizing the magnetization of the free layer and driving it into auto-oscillation with a free-running frequency $f_0$. Additionally, an AC current $I_\textrm{ac}$ generated by an RF source is injected into the STNO. When the frequency of $I_\textrm{ac}$ is close to integer or certain fractional multiples of the STNO frequencies ($f_\textrm{ext}\approx nf_0$) with a sufficiently large amplitude, synchronization occurs between the STNO's oscillations and $I_\textrm{ac}$~\cite{Urazhdin2010Fractional}.

We focus on the $n=2$ case; however, our methods can be easily generalized to integer $n>2$. At zero temperature for the case of $n=2$ when the STNO synchronizes precisely to double its free-running frequency, the phase difference between the oscillator and the external driving force $\Delta\phi = \phi_\textrm{STNO}-\phi_\textrm{ext}$ has only two stable configurations corresponding to in-phase $\Delta\phi\equiv 0$ and out-of-phase $\Delta\phi\equiv \pi$ oscillations, where the phase differences are defined modulo $2\pi$ and modulo an additional constant phase shift, that is neglected here for simplicity \cite{litvinenko2021analog}. At finite temperature, random fluctuations caused by thermal noise excite flucuations in the free-layer magnetization of STNOs. These fluctuations cause the STNO to transition between the two equivalent phase configurations, resulting in a stochastic switching of the phase difference $\Delta\phi$.

For the present application, the important property of the system is the equal probability of finding the STNO with either of the two metastable phases relative to the injected AC signal. We consider an STNO that is injection-locked by an external drive with frequency $\approx 2{f_0}$, i.e. close to twice the free-running frequency. When the phase of the oscillator changes through $2\pi$, it passes through two equivalent phases of the injected external drive, see Fig.~\ref{fig1}. As far as the dynamics of the STNO phase is concerned, the two equivalent phases are symmetric. The symmetry of the dynamics gives rise to the equal probability of being close to either of these phase relationships. This symmetry is discussed in more detail in Sec.~\ref{sec:three}.


We specifically concentrate on STNOs with a magnetic vortex in the free layer because their properties are well established and understood and they have relatively large emission power \cite{tsunegi2016microwave}. However, our approach can be easily applied to STNOs with nominally uniform magnetizations \cite{bonetti_spin_2009, litvinenko2022ultrafast}. Additional details on the vortex-based STNOs used in experiments are provided in Appendix~\ref{appendixA} and~\ref{appendixB}. 

\begin{figure}
 \centering
  \includegraphics[width=86mm\centering]{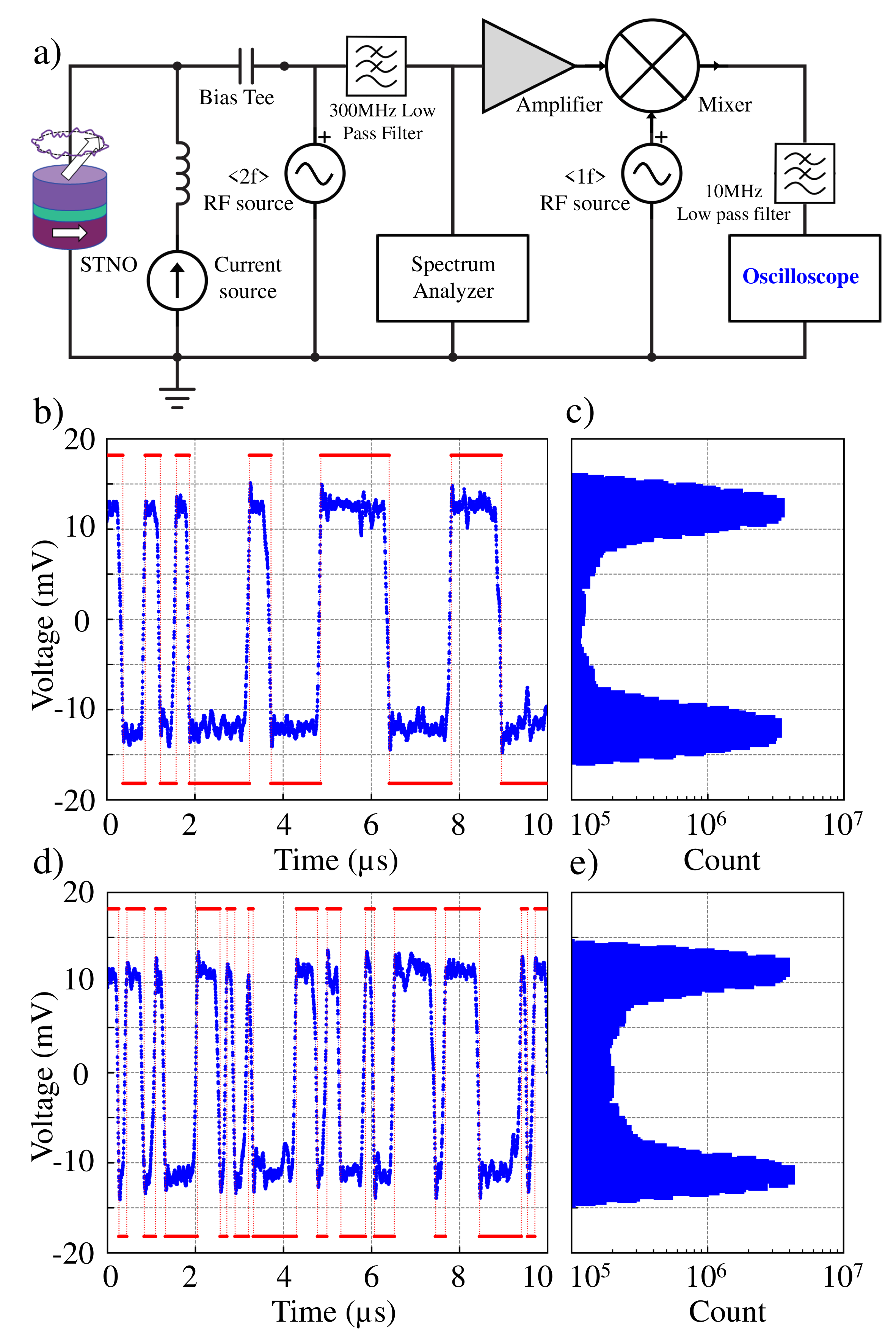}
  \caption{ a) Experimental circuit schematic. A DC current source is used to inject a constant current into the STNO inducing auto-oscillations at $f_0$ frequency. Injection-locking at $2f_0$ is achieved using an RF signal ($f_\textrm{ext}\approx 2f_0$) via a bias tee. The resulting AC signal is low-pass filtered at 300~MHz for STNO dynamics analysis. A phase readout stage, including an operational amplifier and RF mixer, captures phase information. A 10~MHz low-pass filter eliminates rapid variations, enabling phase fluctuation observation on an oscilloscope.  b) The blue graph represents a portion of the voltage-time trace captured for $f_\textrm{ext} = 468$~MHz external frequency and 1~mW power amplitude, spanning a duration of 10~{\textmu}s and exhibiting approximately 10 voltage transitions. The distinct noise levels observed around the two voltage states indicate a symmetric configuration. Both the lower and higher voltage states are equally stabilized, resulting in comparable thermal fluctuations. The red trace represents the binarized version of the blue signal, excluding incomplete transitions or local noise around the stable states.  c) Histogram displaying measured voltages throughout 0.2~s, encompassing approximately $2.5 \times 10^{5}$ transitions. The x-axis is logarithmic, visually suppressing the count rate in the peaks as compared to that in between them. The y-axis is shared with panel (b) d). Graphs equivalent to panel (b) but obtained for $f_\textrm{ext} = 470$~MHz external frequency. The voltage-time trace exhibits faster fluctuations compared to those in panel (b). However, the corresponding histogram e) still shows similar noise levels and equal probabilities for both phase states.}
  \label{fig2}
\end{figure}

To exploit the symmetry of phase fluctuations in STNOs when injection-locked at $2f_0$, we have developed a straightforward electrical phase readout circuit. Figure~\ref{fig2}(a) illustrates the electrical schematic of the circuit, designed not only to measure the conventional RF dynamics of the STNO but also to enable efficient phase detection between the STNO and the injected RF signal. We utilize a constant DC current source that injects a current of $I_\textrm{dc}=7$~mA into the STNO, inducing magnetoresistance auto-oscillations at a frequency of $f_0=229.1$~MHz in the particular STNO. To injection-lock the oscillator at twice the STNO frequency, we introduce an AC signal generated by an RF source referred to as $f_\textrm{ext} \approx 2f_0$. We use a bias-tee to separate AC from DC components. The overall AC signal passes through a 300~MHz low-pass filter, effectively eliminating the injected AC signal's main harmonic and allowing for collection of the RF signal originating from the STNO dynamics. 

The resulting AC signal is split into two parts; one directed toward a spectrum analyzer for frequency domain analysis and the other directed to a designed phase readout stage. This stage includes an operational amplifier with an effective gain of 42~dB, which is primarily used to prevent any back reflection from the phase readout circuit that could alter the STNO dynamics. We should notice that the RF emission amplitude of the vortex-based STNO is sufficiently large in our experiments, allowing the phase readout circuit to function without the need for amplification. The output of the amplifier is further processed by an RF mixer, where it is multiplied by an AC signal at half the frequency of the RF source, i.e., ${f_\textrm{ext}}/{2}$, which coincides with the STNO's frequency in the injection-locking regime. One should notice that instead of using an RF mixer, alternative techniques based on diodes and RF combiners \cite{koo2020distance} can be applied for the same purpose. The output of the mixer can be decomposed into two harmonic signals: a rapidly varying harmonic $\propto \cos (\pi f_\textrm{ext}t+\Delta \phi)$ and a slowly varying harmonic $\propto \cos (\Delta \phi)$. Our main interest lies in the second term, which contains exclusively the phase difference information. We employ a low-pass filter with a cut-off frequency below ${f_\textrm{ext}}/{2}$ to filter out the rapidly varying harmonic. In our case, a 10~MHz low-pass filter was employed. Following this filtering, we collect the resulting signal on an oscilloscope, where the phase fluctuations are contained in a voltage term $\propto \cos(\Delta \phi)$. With this down-conversion method and carefully designed phase readout circuit, we can accurately access and analyze the phase dynamics of the STNO.

The voltage-time trace signal $\propto \cos(\Delta \phi)$ in Fig.~\ref{fig2}(b) shows the stochastic phase dynamic behavior of the system. In order to evaluate the mean dwell times in each phase state and the associated occupation probabilities of those states, we recorded and analyzed from $10^{5}$ to $3\times10^{5}$  transitions. There is an exponential decay in voltage during the transition between the two states, corresponding to an effective $RC$ time constant of 46.6~ns. We attribute this to the combined circuit bandwidth limitations caused by the oscilloscope, low-pass filter, cable capacitance, and the output of the mixer. Our inability to observe dwell times shorter than the characteristic $RC$ decay can have a significant impact on the apparent mean dwell times of the devices. However, for dwell times exceeding approximately 200~ns, the influence of $RC$ dynamics becomes negligible, allowing an accurate characterization of the system's behavior. By employing a robust binning technique, we obtain probabilities associated with each state, as demonstrated in Fig.~\ref{fig2}(c). Importantly, these probabilities remain independent of the specific time durations allocated to each state, providing a reliable measure of state probabilities.

To accurately determine the mean dwell time for each state, we incorporate a numerical digitization technique that is detailed in Appendix~\ref{appendixC}, which excludes rapid transitions over the barrier and back that are incomplete or unsuccessful attempts to transition from one state to another. The resulting refined digital signal is visually depicted in red in Fig.~\ref{fig2}(b), providing a clearer representation of the system dynamics. The evaluation of mean dwell times entails the identification of inter-transition intervals within the refined digital signal. Specifically, we focus on extracting the corresponding dwell times for the high-voltage and low-voltage states observed in the digital (red) signal. These dwell times represent the duration of stability in each state before subsequent transitions occur. We fit the cumulative probability density function of these extracted dwell times with an exponential distribution (see Appendix~\ref{appendixC}) to understand the statistical behavior of the system. This fitting process captures the underlying statistical properties of the dwell times characterized by the mean dwell time as the sole fitting parameter of the exponential distribution. The inverse of the sum of the extracted two mean dwell times is the bit flip rate of this system.

\section{Tunable bit flip rate for Unbiased Random Number Generation}
\label{sec:three}

\begin{figure}
	\includegraphics[width=85mm]{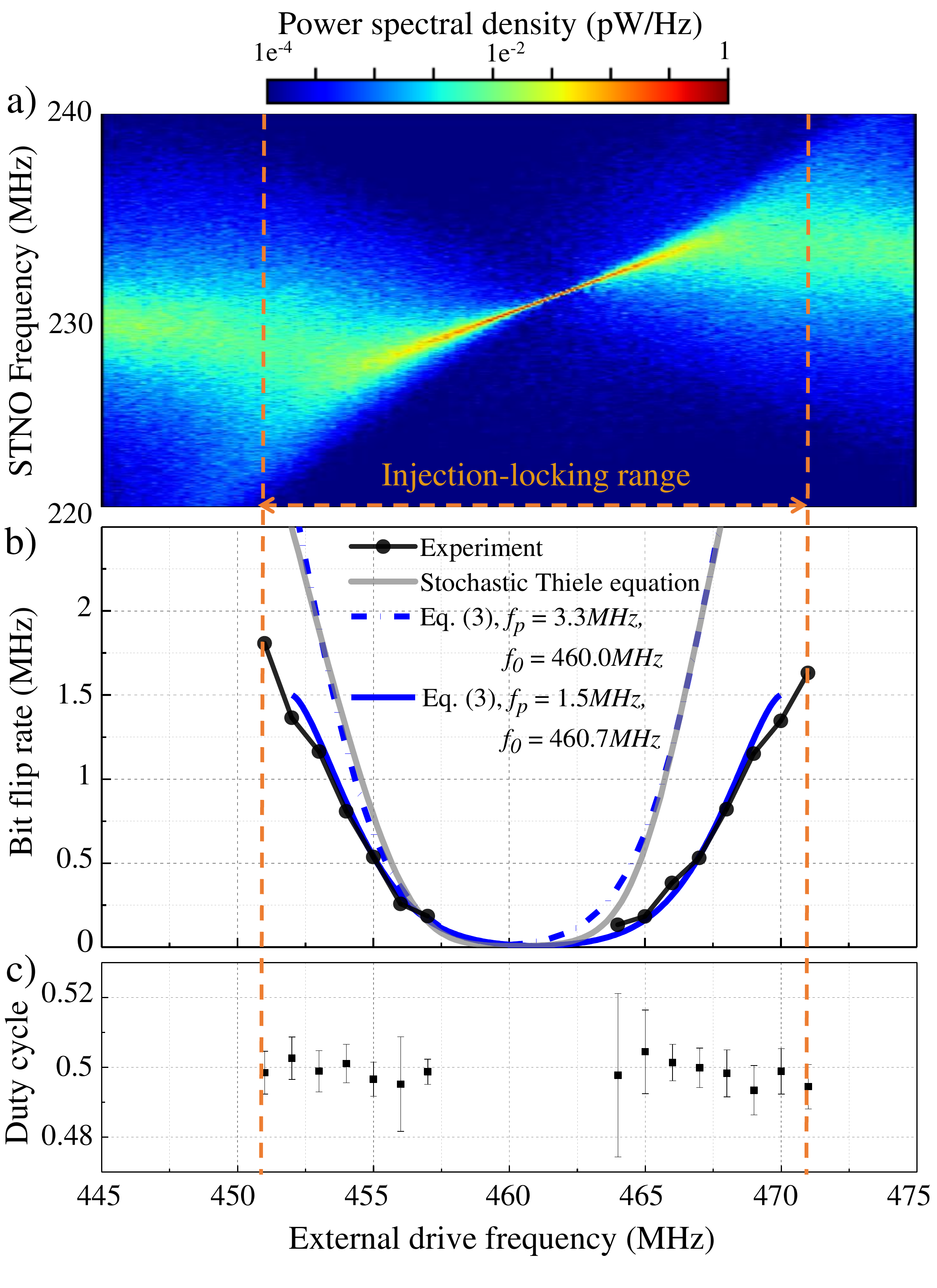}
	\caption{ a) Power spectral density of the microwave emissions of a spin-torque nano-oscillator under an external signal. Frequency sweeps from 445~MHz to 475~MHz, with the injection locking range from 451~MHz to 471~MHz. b) Mean bit flip rate of the phase fluctuations as a function of the external drive frequency, showing the relationship between external drive frequency and bit flip occurrence. The black solid line corresponds to the bit flip rate evaluated in experiments. The solid gray line represents the bit flip rate evaluation obtained from the numerical simulation of the stochastic Thiele equation approach (see Appendix~\ref{appendixD}). The solid and dashed blue lines represent the analytical evolution of the bit flip rate according to Eq.~\ref{Eq3} for two distinct sets of parameters, as indicated in the legend. These parameters respectively capture the trends of the experimental (black) and numerical (gray) results. c) Duty cycle of the phase fluctuations between the two states in experiments, indicating an equal unbiased probability. Error bars were extracted for a 95~\% confidence interval from the mean dwell time evaluation (see Appendix~\ref{appendixC}).}
  \label{fig3}
\end{figure}

Through an analysis of the voltage-time trace recorded at various external frequencies within the synchronization bandwidth, we extract the mean dwell times for the two possible phase states. With those results, we evaluate the bit flip rate as we vary the external source frequency $f_\textrm{ext}$, covering the injection-locking range of the STNO. Figure~\ref{fig3}(a) illustrates the power spectral density of the STNO's microwave emissions under the influence of the external 2$f_\textrm{ext}$ drive signal at a specific operating point ($I_\textrm{dc}=7$ mA, $\mu_0H=300$ mT), and a fixed external drive power of 0.22~mW received by the STNO. When the difference between the STNO's frequency and half of the drive signal's frequency is minimal, the STNO deviates from its natural frequency $f_0$ and closely tracks half of the drive's frequency within the synchronization range. As anticipated, the power spectral density displays enhanced coherence in the central region of the injection-locking range while it diminishes towards the boundaries. 

We record voltage-time traces similar to the one depicted in Fig.~\ref{fig2}(b,d) at various external $f_\textrm{ext}$ frequencies. Outside the locking-range region, no discernible binary-level signal is observed, indicating the unhindered evolution of the STNO phase, consistent with an unsynchronized oscillator. Figure~\ref{fig3}(b) showcases the changing bit flip rate due to the stochastic fluctuations in the STNO's phase, as a function of the external drive frequency. The bit flip rate reaches its peak at the boundaries of the injection-locking range, approaching 2.1~MHz at a 450~MHz external drive frequency, while within the central region of the locking range (459~MHz to 464~MHz), it drastically diminishes to sub-kilohertz fluctuations.

 \begin{figure}
	\includegraphics[width=85mm]{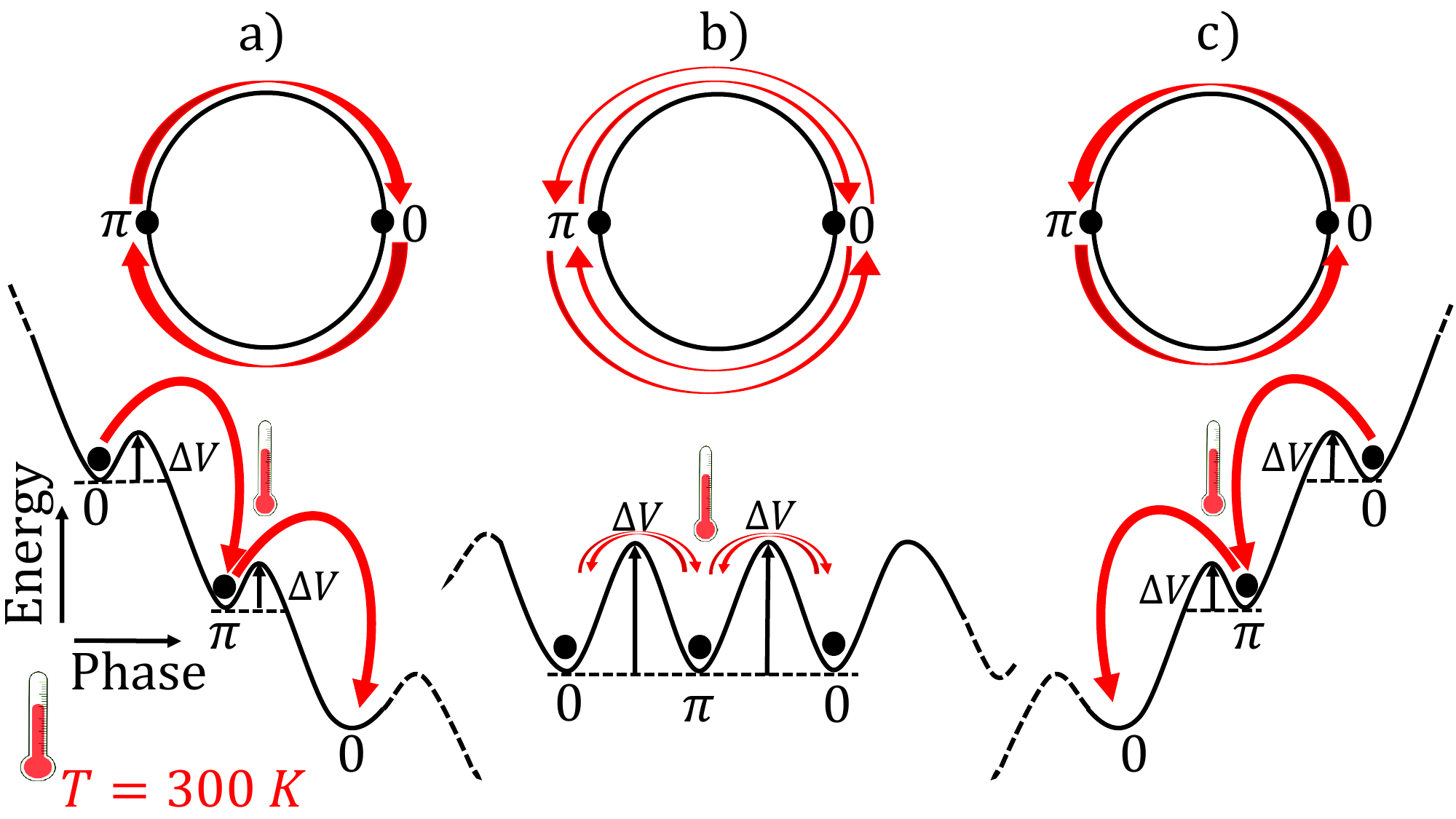}
	\caption{Effective energies of the phase potential. a) Negative frequency detuning with preferential transitions toward positive phase values (clockwise in the figure). Red arrows indicate thermally activated transitions between 0 and $\pi$ states. b) Zero frequency detuning with a sinusoidal energy landscape with a larger barrier between 0 and $\pi$ states. c) Positive frequency detuning with preferential transitions in a negative phase direction (counterclockwise in the figure).}
  \label{fig4}
\end{figure}

In all recorded voltage-time traces, the mean dwell times at a fixed drive frequency, for both phase states are found to be the same within statistical uncertainties. To quantify this relationship, we compute the duty cycle as the ratio of the mean dwell time of the top voltage level to the sum of the two mean dwell times. Figure~\ref{fig3}(c) illustrates the evolution of the duty cycle as a function of the external drive frequency, revealing a consistent value of 0.5. This observation suggests an equal probability of occupying both of the two states, indicating an absence of probability bias throughout the entire injection-locking range.

Despite the inherent stochasticity of the measured phase fluctuations, Fig.~\ref{fig3} demonstrates an elegant approach to deterministically control the bit flip rate while maintaining unbiased probabilities. The tunability of the bit flip rate covers a broad range of timescales, ranging from 1~kHz to almost 2~MHz. In the experiment, this control is achieved by adjusting the external drive frequency within the range of 451~MHz to 456~MHz. A similar dependence for the bit flip rate with unbiased probabilities can be achieved by modulating the power of the external drive. However, in this study, we specifically focus on the frequency dependency for simplicity and phase readout clarity.

Our experimental observations have been rigorously verified through extensive numerical simulations based on the Thiele equation approach subjected to thermal fluctuations. Additional details are presented in Appendix~\ref{appendixD}. The simulations provide strong support for our experimental observations. Figure~\ref{fig3}(b) compares the simulation results (gray line) with the measured bit flip rate (black curve). In spite of the oversimplification of the Thiele-model, see App.~\ref{appendixD}, its simulations qualitatively capture the evolution of the bit flip rate as a function of the external drive frequency, reinforcing the interpretation of our experimental measurements. Additional simulations with this model confirm the unbiased probabilities throughout the entire injection-locking range.

To gain insight into the variation of the bit flip rate as a function of the external drive frequency within the injection-locking range, it is helpful to examine the phase evolution of the system in the effective potential energy landscape created by the injected AC drive. After integrating the phase rate equation of an injection-locked oscillator which is similar to an Adler equation \cite{slavin_nonlinear_2009}, we can express the effective potential energy landscape of the system that depends on the phase difference $\Delta \phi$ as
\begin{align}
\label{eq1}
 V(\Delta \phi) = \delta \Delta \phi - \frac{\Omega}{2} \cos(2\Delta \phi - \phi_0).  
\end{align} 
where $\delta ={f_\textrm{ext}}/{2}-f_0$ denotes the frequency detuning between the external drive and the STNO, $\Omega$ represents the STNO's injection-locking range that is proportional to the coupling strength and the drive power, and $\phi_0$ is a constant phase difference. 
After determining the stable and unstable phase difference states, one can compute the difference between the potential of those minima and maxima, which can be interpreted as a pseudo-energy barrier. When the frequency detuning is non-zero as shown in Fig.~\ref{fig4}(a,c), there are two potential difference heights, one larger and one smaller than the zero-detuning barriers illustrated in Fig.~\ref{fig4}(b). In this analysis, we assume that the phase jump processes are governed by the smaller potential difference height, which can be expressed as 
\begin{align}
\label{Eq2}
\Delta V =  \left( \sin^{-1}\left(\frac{\delta}{\Omega}\right) -\text{sign}(\delta) \frac{\pi}{2} \right) \delta +\Omega \sqrt{1-\left(\frac{\delta}{\Omega }\right)^2}.
\end{align} 
By applying thermally activated escape rate theory and assuming that the system is at equilibrium with the thermal bath and neglecting the curvature of the potential, the bit flip rate corresponding to the transition rate to surmount $\Delta V$ in the presence of thermal energy $kT$, where $T$ is room temperature, and $k$ is the Boltzmann constant, follows an Arrhenius-type law,
\begin{align}
\label{Eq3}
\omega_{\text{bit flip}}\approx  f_p \exp\left(-\frac{\Delta V}{\gamma_0 kT}\right).
\end{align} 
Here $\gamma_0$ is a scaling factor that relates the amplitude of the phase fluctuations to the thermal energy that depends on the damping and the STNO's amplitude radius of its oscillations, and $f_p$ is the relaxation frequency of the oscillator \cite{grimaldi2014response}, (see parameters in Appendix~\ref{appendixD}, Table~\ref{tab:Parameters eq3}). Using this phenomenological model (Eq.~\ref{Eq3}), we were able to achieve agreement between the evolution of the experimental bit flip rate (black curve) and the model (solid blue line) in Fig~\ref{fig3}(b). Similar agreement is seen between the simulated bit flip rate (gray line) and the same model (dashed blue line). The specific parameters used for the comparison between the experiments and simulations in our study are detailed in Appendix~\ref{appendixD} in Table~\ref{tab:Parameters eq3}.

This simple phenomenological model uncovers two regimes that warrant further examination. The first regime pertains to a small effective energy barrier, characterized by $\Delta V \ll\gamma_0 kT$. In this regime, transitions primarily give rise to the linear dependence on the frequency detuning of the bit flip rate $\omega_{\text{bit flip}} \propto  f_p(1-{|\Omega-\delta|}/{\gamma_0 kT})$. This regime aligns with the observed linear dependency of the bit flip rate on the external drive frequency, as depicted in Fig.~\ref{fig3}(b) for two frequency windows: 450~MHz to 455~MHz and 465~MHz to 471~MHz. These windows correspond to negative and positive frequency detuning values, respectively. Qualitative illustrations in Fig.~\ref{fig4}(a) and (c) depict the most probable thermally activated transitions occurring in a periodically modulated energy potential influenced by the frequency detuning in this considered regime.

On the other hand, the second regime corresponds to larger effective energy barriers, where $\Delta V \gg\gamma_0 kT$. This regime gives the observed nonlinear dependency of the bit flip rate on the external drive frequency, as observed in Fig.~\ref{fig3}(b) for the central frequency window 455~MHz to 465~MHz, associated with smaller frequency detuning values. Fig.~\ref{fig4}(b) provides a qualitative depiction of the most probable thermally activated transitions occurring in a periodic energy potential that remains untilted by the frequency detuning contribution.

When the frequency detuning approaches zero ($\delta=0$), corresponding to $f_\textrm{ext}={f_0}/{2}$, the effective energy barrier reaches its maximum value. Consequently, the transition rate is minimized due to the exponential dependence of $\omega_{\text{bit flip}}$ on the energy barrier. In contrast, when the system operates near the boundary of the injection-locking range ($\delta=\Omega$), the effective energy barrier is minimized, resulting in the highest rates of bit flip transitions possible in the system.

\section{Unbiased Generation of Truly Random Numbers and Benchmarking}
\label{section IV}

To evaluate the randomness of the generated bitstream, we use the National Institute of Standards and Technology Statistical Test Suite (NIST STS)\cite{bassham2010sp}. This suite applies 188 statistical tests to assess the randomness of bitstreams. These tests analyze various statistical properties to determine if the bitstream exhibits characteristics consistent with perfect randomness, such as mean value, autocorrelation, standard deviation, estimated entropy, and pattern occurrence frequencies. Before evaluation, the analog signal captured by the oscilloscope must be digitized into a binary bitstream of ones and zeros. We choose to digitize the signal numerically on a computer, however, the digitization process can be easily achieved electrically using a comparator-based circuit. Given that the analog signal is centered around a zero voltage level, i.e. is unbiased, we designated positive analog values as bit 1 and negative voltages as bit 0. In other words, the digitization voltage threshold is set to exactly zero.

To test the randomness, a bitstream of $10^8$ bits was divided into one hundred $10^6$-bit sequences. Each sequence underwent independent testing, and the pass rate (percentage of $10^6$-bit sequences passing the test) was computed for all 188 tests. The results of these tests, conducted at an external drive frequency of $f_\textrm{ext}=468$~MHz, are summarized in Table \ref{table:1}. Operating at this drive frequency allows for a sufficient number of transitions within a reasonable memory size that the oscilloscope can allocate for recording the corresponding voltage-time trace. The bitstream successfully passed all 188 tests, indicating that it is consistent with perfect randomness. This outcome demonstrates the robustness and quality of the generated bitstream in terms of its statistical properties, providing confidence in its suitability for various applications requiring secure and reliable random data generation.

\begin{table}[]
\centering
\caption{
NIST Statistical Test Suite results on a $10^{8}$ bit stochastic bitstream obtained at an external driving frequency $f_\textrm{ext}=468$~MHz and a sampling frequency $f_\textrm{s}=500$~kHz. }
\begin{tabular}{|l|c|c|}
\hline
\textbf{Statistical Test} & \multicolumn{1}{c|}{\textbf{Pass rate}} & \textbf{Status} \\ \hline
Frequency                 & 0.99                                     & Pass            \\ \hline
BlockFrequency            & 0.97                                     & Pass            \\ \hline
CumulativeSums (Forward)  & 0.99                                     & Pass            \\ \hline
CumulativeSums (Reverse)  & 0.98                                     & Pass            \\ \hline
Runs                      & 0.98                                     & Pass            \\ \hline
LongestRun                & 0.99                                     & Pass            \\ \hline
Rank                      & 1.00& Pass            \\ \hline
FFT                       & 0.99                                     & Pass            \\ \hline
NonOverlappingTemplate    & 0.99                                     & Pass            \\ \hline
OverlappingTemplate       & 1.00                                        & Pass            \\ \hline
Universal                 & 0.98                                     & Pass            \\ \hline
ApproximateEntropy        & 0.99                                     & Pass            \\ \hline
RandomExcursions          & 0.98                                     & Pass            \\ \hline
RandomExcursionsVariant   & 0.98                                     & Pass            \\ \hline
Serial                    & 0.99                                     & Pass            \\ \hline
LinearComplexity          & 0.99                                     & Pass            \\ \hline
\end{tabular}

\label{table:1}
\end{table}

 \begin{figure}
  \centering
	\includegraphics[width=85mm]{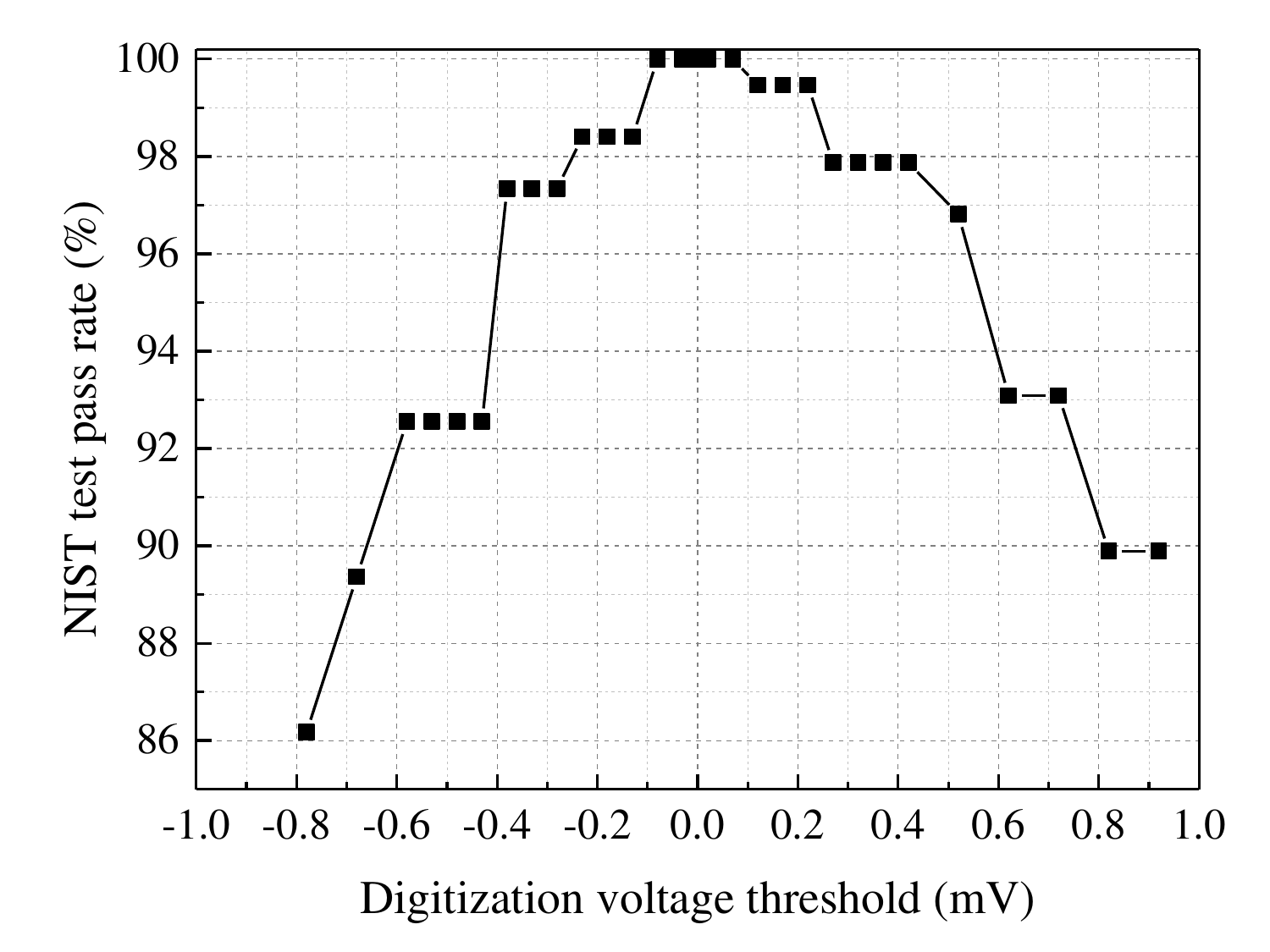}
	\caption{ Evolution of the NIST test pass rate evaluated for different values of the digitization voltage threshold. The NIST test results depicted in Table \ref{table:1} correspond to the zero voltage threshold. }
  \label{nisttest}
\end{figure}

To investigate the potential limitations of our random number generation method, we study scenarios where the digitization process might compromise the bitstream's randomness. Specifically, we examine situations where the voltage threshold for digitizing the analog signal deviates slightly from zero. Such deviations might arise in mildly asymmetric digital circuits or those prone to pronounced device-to-device variability. In Fig.~\ref{nisttest}, we adjust the digitization voltage threshold from $-1.0$~mV to 1.0~mV on the x-axis and evaluate the randomness of the resultant bitstream using the NIST test. Predictably, the pass rate drops from 100~\% to nearly 86~\%. This reduction is attributed to an increasing imbalance between the counts of zeros and ones in the digital bitstream. Yet, within a 150~{\textmu}V span centered on the zero-voltage threshold, the pass rate remains a consistent 100~\%. This suggests our method displays resilience against minor unwanted voltage offsets introduced during digitization.

\section{Conclusion}

Our exploration of stochastic phase dynamics in STNOs has focused on harnessing their intrinsic noise, shifting away from traditional phase coherence improvement efforts. By injecting an AC current close to integer multiples of the STNO's natural frequency, we observed synchronization, notably at double the frequency, leading to two stable phase configurations. Thermal noise at finite temperatures induces random transitions between these configurations, producing a stochastic switching of the phase difference. A highlight is a down-conversion technique using a phase readout circuit presented in Sec.~\ref{section II}. This circuit discerns the STNO's phase dynamics efficiently, providing an accurate representation of its inherent stochastic behavior. Through this, we see the potential of STNOs not just for generating random bitstreams but for broader STNO phase-based applications.

Our work elucidated the influence of varying external drive frequencies on the bit flip rate of the corresponding stochastic bitstream within the synchronization bandwidth of a vortex-based STNO. Key findings in Sec~\ref{sec:three} reveal that the bit flip rate can be deterministically controlled, reaching a maximum of 2.1 MHz bit flip at a 450 MHz external drive frequency. Notably, throughout the various changes in the external drive frequency, the system consistently exhibited a duty cycle value of 0.5, pointing to an absence of a discernable probability bias. These findings hold significant implications for optimizing the operational characteristics of spintronic devices and offer promising avenues for their application in random number generation.

We rigorously evaluated the randomness of our generated bitstream using the NIST STS benchmark. At an external drive frequency of 468MHz, the success pass rate was 100~\% for digitization threshold levels within a 150~{\textmu}V span around zero. However, broader threshold deviations underscored the need for precision, as they introduced potential imbalances. Ultimately, Sec.~\ref{section IV} reaffirms the robustness of our method, its resilience to minor perturbations, and its promise for applications necessitating secure unbiased random number generation.

 To reach higher bit flip rates, we believe our approach can be easily adapted to nonlinear oscillators operating at higher frequencies. For instance, STNOs operating at several dozens of gigahertz \cite{litvinenko2022ultrafast, bonetti_spin_2009} that can synchronize to external stimuli could be used to obtain fast stochastic phase fluctuations in the gigahertz range. Such approaches will be very attractive to explore for fast and unbiased random number generation schemes addressing high-speed encryption applications at the nanoscale.

\section*{Acknowledgements}
This work was supported partially by NSF-ANR via grant StochNet Project ANR-21-CE94-0002-01 and NSF grant number CCF-CISE-ANR-FET-2121957, and partially supported by Grenoble INP Bourse Présidence and MIAI@Grenoble Alpes (ANR-19-P3IA-0). The authors would like to thank Matthew W. Daniels, Louis Hutin, Matthew R. Pufall, and Jabez J. McClelland for fruitful discussions and suggestions.
\appendix
\label{appendix}

\section{Vortex-Based Spin-Torque Nano-Oscillator Structure}
\label{appendixA}
We chose to use vortex-based STNOs in this work. The devices were developed at the International Iberian Nanotechnology Laboratory (INL). The layer stack from which the magnetic tunnel junctions were fabricated has the composition: Ta(5)/CuN(50)/Ta(5)/CuN(50)/Ta(5)/Ru(5)/IrMn(6)\\/CoFe$_{30}$(2.6)/Ru (0.85)/CoFe$_{40}$B$_{20}$(1.8)/MgO/VFL/\\Ta(10)/CuN(30)/Ru(7). Numbers in parentheses represent thicknesses in nanometers. The materials for the magnetic tunnel junction were deposited using a sputter deposition process and nanofabricated at the INL along with ion beam and optical etching techniques. Of particular significance is the VFL, or vortex-state free layer, that corresponds to CoFe$_{40}$B$_{20}$(2.0)/Ta(0.2)/NiFe(7). The results in this work were for devices  400~nm in diameter. The static resistance of the device was around 40~$\Omega$, with a tunnelling magnetoresistance of approximately 150~\% and a resistance-area product close to
4.5~\unit{\ohm\micro\meter\squared}.

\section{Free-Running and Synchronization Dynamics of the Spin-Torque Nano-Oscillator}
\label{appendixB}

\begin{figure}
  \centering
	\includegraphics[width=87mm]{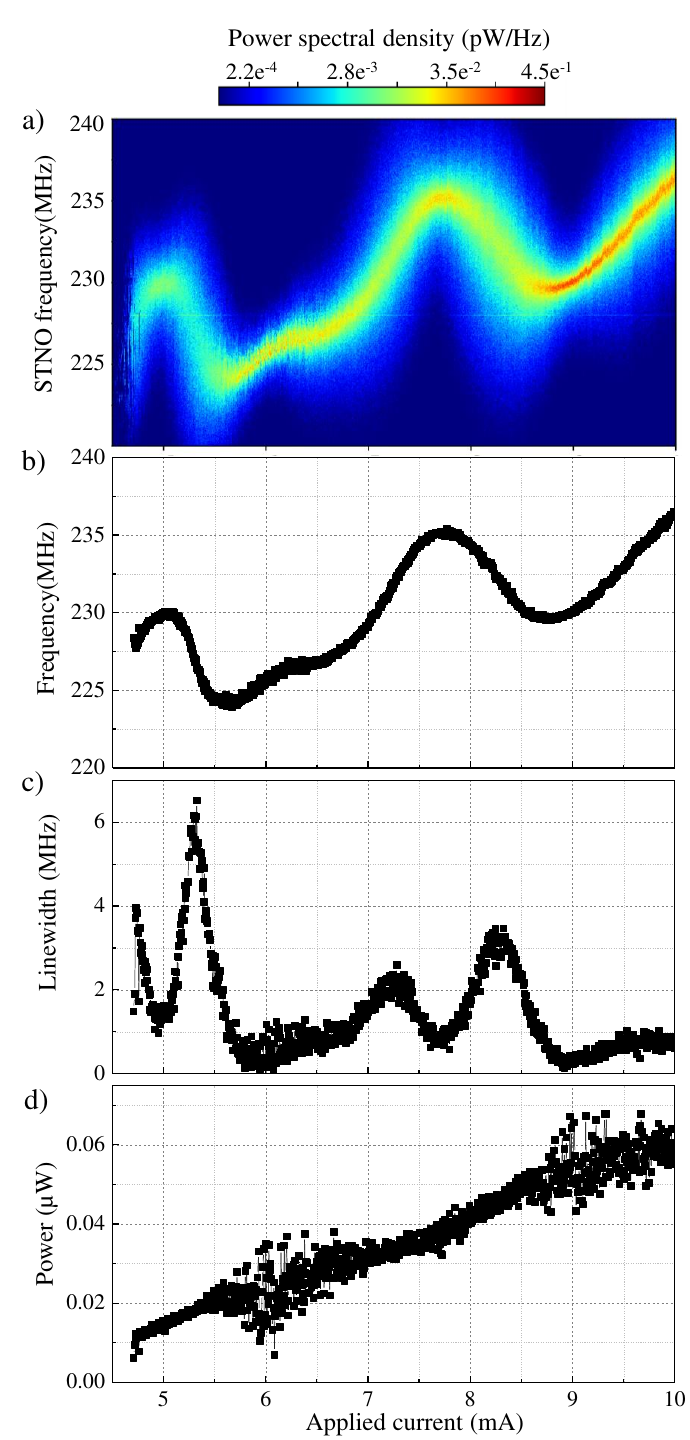}
	\caption{ a) Experimental power spectral density measured for the studied STNO in free-running regime corresponding to the absence of an external microwave drive at a fixed perpendicular applied magnetic field $\mu_0H=300$~mT. The applied current is swept on the x-axis from 4.5~mA to 10~mA. The bright regions illustrate the evolution of the free-running frequency as a function of the applied DC current. b) Evolution of the peak free-running frequency versus applied DC current. The frequency is evaluated from the Lorentzian fit of experimental frequency spectra. c) Corresponding evolution of the linewidth extracted from Lorentzian fits. d) Corresponding integrated power evaluated from the fits. }
  \label{free}
\end{figure}

Figure~\ref{free} illustrates the free-running vortex-based STNO device's frequency-current characteristics. In the range 4.5~mA to 10~mA, the behavior is not as simple as would be expected for an ideal gyrotropic oscillator that would be described by the Thiele model. We choose to operate in all of the measurements in this paper with applied DC current of 7~mA giving a free-running frequency $f_0$ = 230~MHz. This operating point was driven by the large observed frequency tunability that should lead to enhanced ability to injection-locked to external signals.

\begin{figure}
  \centering
	\includegraphics[width=87mm]{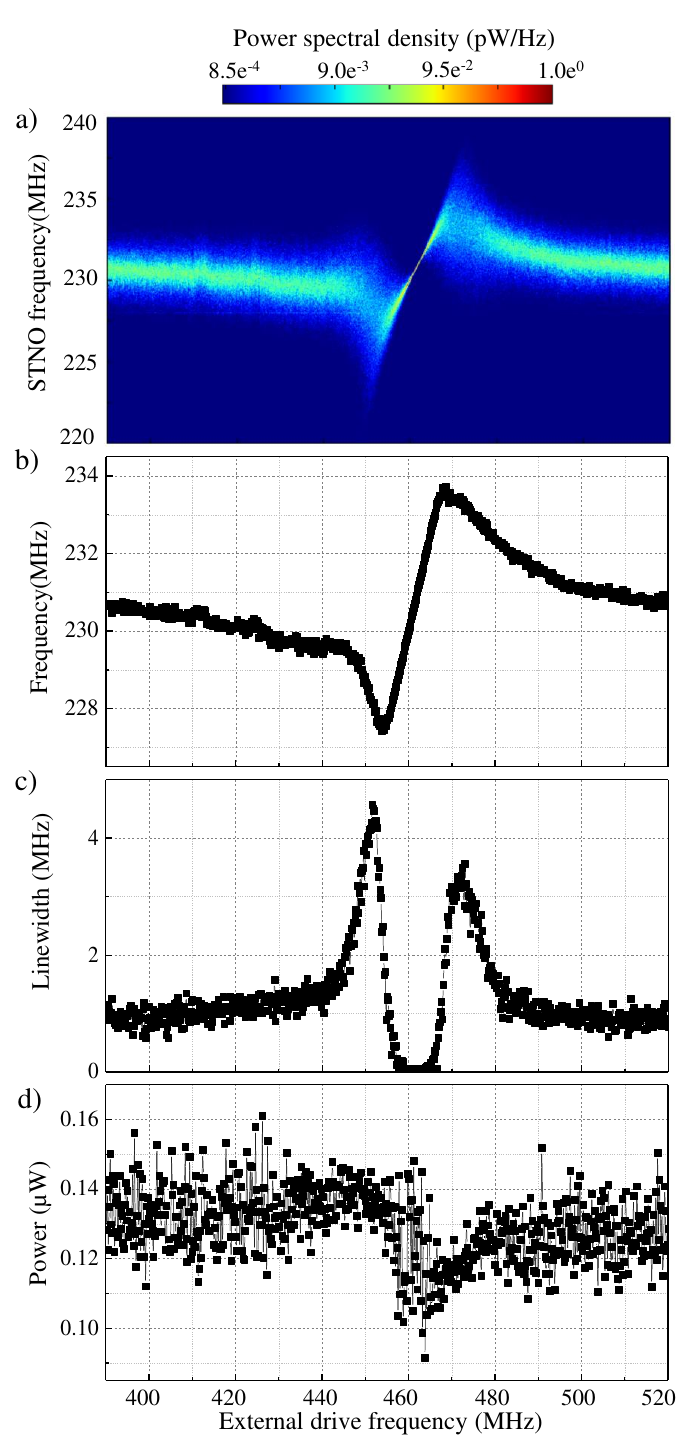}
	\caption{a) Power spectral density corresponding to the microwave emissions of the studied STNO obtained at 7~mA DC applied current and in the presence of an external microwave signal with a frequency swept from 390~MHz to 520~MHz. The injection locking range corresponds to the central area contained between 450~MHz and 470~MHz. The drive amplitude is maintained constant at 0.13~\unit{\micro\watt}. b) Evolution of the peak frequency of the STNO from Lorentzian fits versus external drive frequency. c) Corresponding evolution of the extracted linewidth versus external drive frequency exhibiting an increase and reduction of the linewidth, respectively, at the borders and center of the injection-locking range. d) Evolution of integrated power evaluated from the fits versus the external drive frequency.}
  \label{fig6}
\end{figure}

In order to investigate the injection locking behavior at 2$f_0$, we conducted experiments at the chosen operating point of 7~mA, maintaining a constant current while sweeping the frequency of the external AC drive with a constant RF drive power amplitude  of 0.22~mW received by the STNO. The results, presented in Fig.~\ref{fig6}, reveal the evolution of the oscillator frequency as a function of the drive frequency. As expected, we observed a distinct locking region where the oscillator frequency synchronized with the drive frequency. This behavior indicates injection locking, where the external drive imposes its frequency onto the oscillator. The locked region exhibited a clear and consistent relationship between the oscillator and drive frequencies, confirming the successful injection locking phenomenon.

Figs.~\ref{fig6}(b,c,d) provide additional insights into this injection locking process showing that our experimental results align with the expected injection locking behavior.  The linewidth displays a notable reduction within the locking region, indicating enhanced spectral purity during injection locking. The integrated power graph does not show a significant increase within the locking region, consistent with an interpretation that the integrated powers remain constant even though that power is spread over a narrower frequency window.

\section{Evaluation of the mean bit flip rate}
\label{appendixC}

To assess the bit flip rate, we digitize the analog signal measured by the oscilloscope, using a zero-voltage threshold as our basis. To obtain an accurate estimate of mean dwell times and consequently the corresponding bit flip rate, it is essential to exclude "incomplete" transitions. While various definitions can specify a complete transition, we define it as a scenario where the analog signal consecutively reaches two voltage levels delineating the boundary between high and low states. We opted to identify these levels at the inflection points of the probability distribution function of the entire voltage-time trace signal, which corresponds to the voltage where the derivative of the histogram as a function of the voltage is at its maximum.

\begin{figure}
   \centering
	\includegraphics[width=85mm]{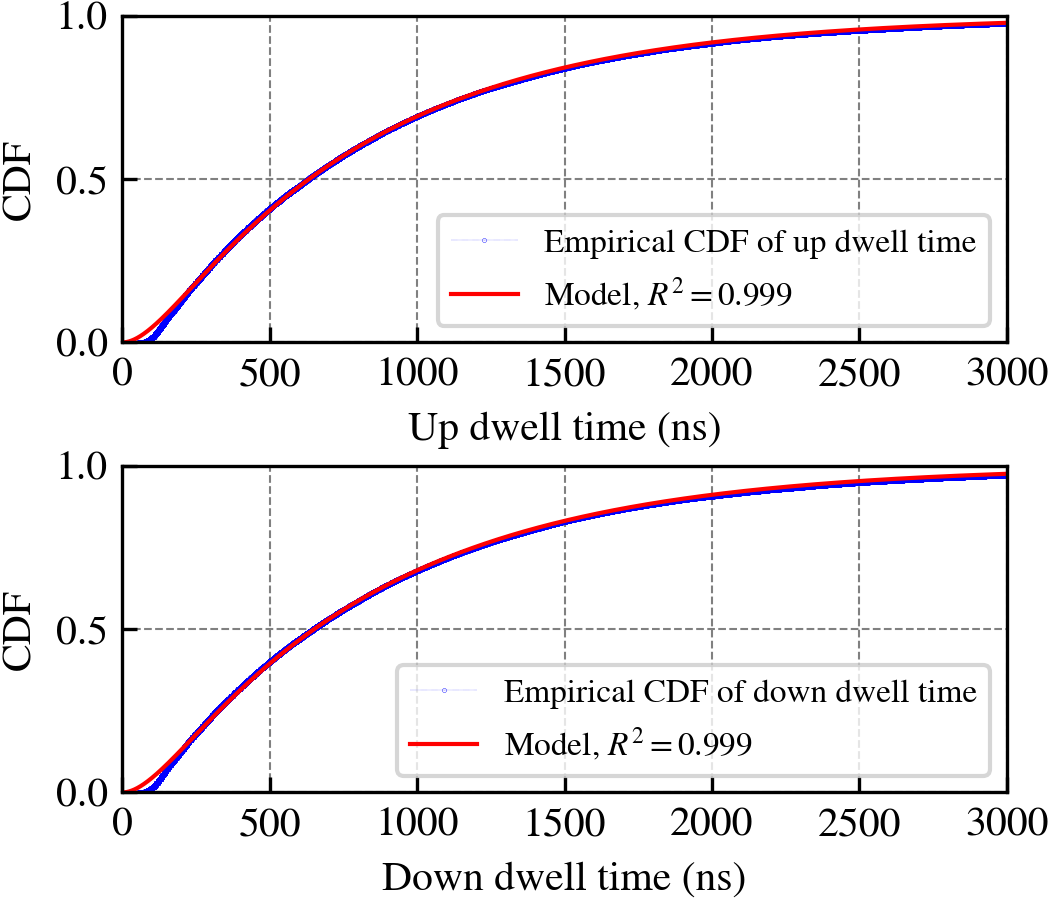}
	\caption{Empirical cumulative distribution function (CDF) of the extracted dwell times for positive and negative voltage levels recorded on the oscilloscope corresponding to the $f_\textrm{ext} =$ 455~MHz external frequency. The red curve shows the corresponding pseudo-exponential distribution CDF convoluted with the finite bandwidth (see Eq.~\ref{eq:CDF-finite}) of measurement process, that was used to fit the empirical data.}
  \label{fig8}
\end{figure}

Post-digitization, we identified the timestamps of every transition between bits, allowing us to determine the dwell time before each bit transition. We subsequently analyzed the statistical distribution of these extracted dwell times by examining the cumulative distribution function (CDF) for both up and down phase states (positive and negative). As depicted in Fig.~\ref{fig8}, the CDF exhibits an almost exponential distribution that is altered by the finite bandwidth of our measurement process. Fitting the data to the exponential distribution is a common practice in analyzing dwell times of experimental random telegraph noise \cite{talatchian2021mutual}. However, this approach showcases a discrepancy for short dwell times. We propose to take into account the impact of the finite bandwidth of the measurement using the 4-state Markov approach developed by Naaman et al.~\cite{naaman2006poisson}. Based on this approach, the empirical CDF, instead of an exponential distribution, is fitted by the following CDF distribution for the 'in state' $A$. In our case, this corresponds to our system's up dwell times (recorded for positive output voltage states)

\begin{equation}
\begin{aligned}
F_{A}(t;\gamma,\Gamma_A,\Gamma_B) = &1 - \frac{e^{-\frac{1}{2} \Gamma_\text{tot}t}}{\mathcal{D}} \mathcal{K} \biggl[\mathcal{K}\cosh\left(\frac{1}{2} \mathcal{K}t\right) \\
&+ \Gamma_\text{tot} \sinh\left(\frac{1}{2} \mathcal{K}t\right) \biggr],
\label{eq:CDF-finite}
\end{aligned}
\end{equation}

where
\begin{equation}
\begin{aligned}
    \Gamma_\text{tot}= \gamma + \Gamma_A + \Gamma_B , \\
\end{aligned}
\end{equation}
\begin{equation}
\begin{aligned}
\mathcal{D} = \sqrt{\gamma^2 + 2 \gamma (\Gamma_A - \Gamma_B) + (\Gamma_A + \Gamma_B)^2},
\label{eq:D def}
\end{aligned}
\end{equation}
\begin{equation}
\begin{aligned}
\mathcal{K}^2 = -4 \gamma \Gamma_A + \Gamma_\text{tot}^2 .
\label{eq:K def}
\end{aligned}
\end{equation}

In this more complex CDF distribution, $\gamma$ corresponds to an effective detection rate corresponding to the physical bandwidth of the measurement process, while $\Gamma_A$ and $\Gamma_B$ correspond, respectively, to the inverse of the mean dwell times in states $A$ and $B$. In our case, these correspond to positive and negative output voltage states. One can compute the CDF for the other mean dwell distribution in state $B$ by exchanging $\Gamma_A$ with $\Gamma_B$ and vice versa. In the limit of infinite measurement bandwidth ($\gamma \to \infty$), we recover the exponential distribution $F_{A(\text{exp})}(t) = 1 - \exp(-\Gamma_{A}t)$.

By utilizing Eq.~\ref{eq:CDF-finite}, we achieved excellent quantitative agreement and determined the mean dwell times for both up and down dwell states (positive and negative). The finite bandwidth of our measurement circuit was fitted and evaluated as $1/\gamma = 46.5$~ns, matching the time scales imposed by the $RC$ dynamics of our circuit. 


The bit flip rate is then derived by taking the inverse of the sum of the two extracted mean dwell times. To provide error bars for the resulting bit flip rate, we incorporated the fitting uncertainty of exponential fits of the dwell time CDFs, set at three times the standard deviation. Those uncertainties are then propagated in the expression of the bit flip rate.A connected quantity that we refer to as the duty cycle, shown in Fig.~\ref{fig3}(c), is computed for each bit flip rate. In our case, this corresponds to the product of the bit flip rate with the mean up dwell time. The uncertainty of this quantity is obtained through error propagation of the mean dwell times, similar to the bit flip rate.


\section{Numerical simulation of the stochastic Thiele equation approach}
\label{appendixD}
\begin{table}[]
\centering
\caption{Spin-torque oscillator parameters used in the numerical simulations.}
\begin{tabular}{lrl}
\hline
\textbf{Symbol} & \textbf{Value}         &                        \\ \hline
$G$   & $2.03 \times 10^{-13}$ & $\mathrm{Js/(m^2 rad)}$ \\
$D_0$  & $4.56 \times 10^{-16}$ & $\mathrm{Js/(m^2 rad)}$ \\
$D_1$  & $ 1.36 \times 10^{-15}$ & $\mathrm{Js/(m^2 rad)}$ \\
$\kappa_\mathrm{MS}^0$ & $2.26 \times 10^{-4}$  & $\mathrm{J/m^2}$       \\
$\kappa_\mathrm{MS}^1$ & $1.00 \times 10^{-5}$  & $\mathrm{J/m^2}$       \\
$\kappa_\mathrm{Oe}^0$ & $2.21 \times 10^{-15}$ & J/A                    \\
$\kappa_\mathrm{Oe}^1$ & $-1.32 \times 10^{-16}$ & J/A                    \\
$a_\mathrm{J}$         & $1.47 \times 10^{-16}$ & J/A                    \\
$b_\mathrm{J}$         &$4.96 \times 10^{-23}$ & J/A                    \\
$r_0$                  & $200$                  & nm       \\
$\lambda$              & 2/3                  &     \\
$\xi$                  & +1                  &        \\\hline           
\end{tabular} \label{tab:ModelParameters}
\end{table}

We model the gyrotropic mode of the vortex spin-torque oscillator using Thiele equations~\cite{Thiele_Steady_1973, khalsa2015critical, Khvalkovskiy_Vortex_2009}. Ignoring vortex distortions, the vortex configuration of the free-layer can be parameterized by the position of the vortex $\textbf{r} = r_0 (\rho \cos\theta,\rho \sin\theta)$, with $r_0$ being the free-layer radius. The time evolution of the oscillator in the presence of fluctuations induced by finite temperature $T$ is described as
\begin{align}
    \dot{\rho} & = a\rho - b\rho^{3} - (c + \eta_x) \cos(\theta) - \eta_y\sin(\theta),  \nonumber \\
    \dot{\theta} & = \omega_0 + \omega_1\rho^2 + \left(\frac{c}{\rho} + \eta_x \right) \sin(\theta)-\eta_y\cos(\theta),\label{eq:ModelBasic}
\end{align}
where
\begin{align}
    a & = \frac{a_\mathrm{J} I}{\pi G r_0^2} - \frac{D_0}{G}\omega_0, \nonumber\\
    b & = \frac{D_1}{G}\omega_0+\frac{D_0}{G}\omega_1, \nonumber\\
    c & = \frac{b_\mathrm{J}I}{\pi G r_0^2}, \nonumber \\
    \omega_0 & = \frac{1}{G}\left(\kappa_\mathrm{MS}^0 + \kappa_\mathrm{Oe}^0 \frac{I}{\pi r_0^2}\right), \nonumber\\
    \omega_1 & = \frac{1}{G}\left(\kappa_\mathrm{MS}^1 + \kappa_\mathrm{Oe}^1 \frac{I}{\pi r_0^2}\right). \label{eq:ModelParameters}
\end{align}
Here, $G$ is the gyrovector amplitude, $D_0$ ($D_1$) is the zeroth (first) order damping constant, $\kappa_\text{MS}^0$ ($\kappa_\text{MS}^1$) is the zeroth (first) order magnetostatic confinement constant, $\kappa_\text{Oe}^0$ ($\kappa_\text{Oe}^1$) is the zeroth (first) order magnetostatic Oersted field confinement constant, $a_\text{J}$ is the orthogonal spin-transfer efficiency parameter, $b_\text{J}$ is the in-plane spin-transfer efficiency parameter, $I$ is the instantaneous current through the spin-torque oscillator, $\eta_x$ and $\eta_y$ are the thermal fluctuation components in Cartesian coordinates at time $t$. The thermal fluctuations are described by $\delta$-correlated Gaussian white noise sources, and thus their ensemble averages are given by
\begin{equation}
\langle\eta_i\rangle = 0,\text{ and }\langle\eta_i(t)\eta_j(t')\rangle = \Gamma\delta_{ij}\delta(t-t'), \label{eq:ensembleNoise}
\end{equation}
where the amplitude of thermal fluctuations $\Gamma = 2k_\text{B}TD_0/(r_0^2G^2)$. We use Itô calculus to obtain the time evolution of the stochastic differential equation \eqref{eq:ModelBasic}.

The measured critical frequency ($\omega_c$), along with the critical current ($I_c$) of the gyrotropic mode are used to extract $M_\text{s}$, and therefore $\kappa_\text{Ms}^0$, $\kappa_\text{Oe}^0$ and $G$ using 
\begin{align}
\omega_c & = \frac{1}{G}\left(\kappa_\mathrm{MS}^0 + \kappa_\mathrm{Oe}^0 \frac{I_c}{\pi r_0^2}\right), \nonumber\\
G & = 2\pi t_0 M_\text{s} \gamma \nonumber\\
\kappa_\mathrm{MS}^0 & = \frac{10}{9} \mu_0 M_\text{s}^2 t_0^2 / r_0, \nonumber\\
\kappa_\mathrm{Oe}^0 & = 0.85 \mu_0 M_\text{s} t_0 r_0.
\end{align}
The ratio $D_0/a_\text{J}$ is also extracted from the critical current by setting $\dot{\rho} = 0$ in Eq.~\eqref{eq:ModelBasic}, and neglecting noise and first order terms, giving $D_0/a_\mathrm{J} = I_c/(\pi r_0^2 \omega_c)$. The remaining first order parameters ($D_1$, $\kappa_\mathrm{MS}^1$, and $\kappa_\mathrm{Oe}^1$) along with $b_\text{J}$ are estimated by matching to measured values the fundamental oscillating frequencies predicted by the model and their DC drive current dependence for the gyrotropic mode of oscillation. 

The fitting procedure described above is overparametrized and there are multiple sets of parameters that can describe the measured oscillator output. The set of parameters that we chose for our numerical model is given in Table \ref{tab:ModelParameters}.

The resistance of the spin-torque oscillator as a function of the oscillating core coordinates is
\begin{equation}
    R = \frac{1}{2}(R_\mathrm{AP}+R_\mathrm{P}) + \frac{\lambda \xi}{2} (R_\mathrm{AP}-R_\mathrm{P}) \rho \sin(\theta), 
\end{equation}
where $\lambda$ is the average magnetization to vortex displacement ratio, $\xi=\pm1$ depending on the helicity and the direction of oscillation of the vortex core. $R_\mathrm{AP}$ and $R_\mathrm{P}$ are the resistances of the STNO when the free layer is magnetized entirely antiparallel and parallel, respectively, with respect to the fixed layer.

The output of voltage across the oscillator is computed by using Ohm's law under the assumption that the vortex oscillator is a purely resistive element at all frequencies. The simulated voltage traces are then passed through a digital filter model that mimics the phase readout circuitry shown in Fig.~\ref{fig2}(a) to obtain the bistable voltage traces similar to the ones in Figs.~\ref{fig2}(b) and (c).


The simple thermally activated model given in Eq.~(\ref{Eq2}) and Eq.~(\ref{Eq3}) can capture the behavior found in both experiment and the Thiele model simulations above. The parameters used to fit those results are in Table~\ref{tab:Parameters eq3}. 
\begin{table}[h]
\centering
\caption{Parameters used to fit experimental and numerical bit flip rate results using Eq.~\ref{Eq3} model in Fig.~\ref{fig3}(b).}
\begin{tabular}{lrl|lrl}
\hline
\multicolumn{3}{c|}{\textbf{To fit experiments}} & \multicolumn{3}{c}{\textbf{To fit simulations}} \\
\textbf{Symbol} & \textbf{Value} & & \textbf{Symbol} & \textbf{Value} & \\ \hline
$fp$   & $1.5$ MHz & & $fp$   & $3.3$ MHz & \\
$f_0$  & $460.7$ MHz & & $f_0$ & $460.0$ MHz & \\
$T$    & $300$ K & & $T$ & $300$ K & \\
$\gamma_0$ & $5.53 \times 10^{23}$~Hz/J & & $\gamma_0$ & $5.53 \times 10^{23}$~Hz/J & \\
$\Omega$ & $10$ MHz & & $\Omega$ & $10$ MHz & \\ \hline
\end{tabular}
\label{tab:Parameters eq3}
\end{table}
We should notice that $\gamma_0$ which is a fit parameter in our approach can be defined as the damping-like torque efficiency~\cite{grimaldi2014response} as
\begin{equation}
\gamma_0  = \frac{1}{2\pi r_0^2 p_0} \left( \frac{D_0}{G^2} \right) \left( 1 + (\frac{k_{\text{MS}}^{1} + k_{\text{Oe}}^{1}(\frac{I}{\pi r_0^2})}{k_{\text{MS}}^{0} + k_{\text{Oe}}^{0}(\frac{I}{\pi r_0^2})} + \frac{D_1}{D_0}) p_0 \right),
\end{equation}
where $p_0$ is the power amplitude of the STNO.



%


\end{document}